\documentclass{article}
\usepackage{amsmath,amssymb,graphicx,a4wide}
\newcommand{\esp}[1]{\mathbb{E} \left( #1 \right)} 

\newcommand{\ind}[1]{\mathbf{1}_{\left\{ #1 \right\}}}

\def\stackunder#1#2{\mathrel{\mathop{#2}\limits_{#1}}} 
\newcommand{\DA}[3]{\stackunder{#1 \rightarrow #2}{#3}}

\newcommand{\tendvers}[2]{\DA{#1}{#2}{\longrightarrow}}

\newcommand{\ordinal}[1]{$#1^{\textrm{th}}$}

\newcommand{\important}[1]{\textbf{#1}}

\newcommand{\gui}[1]{"{#1}"}

\newlength{\tailleimage}
\begin{document}


\title{Estimating copula measure using ranks and subsampling: a simulation study}
\author{J\'er\^ome Collet, \'Electricit\'e de France R\&D division
\footnote{Jerome dot Collet at edf dot fr or Jerome dot Collet at ensae dot org}}

\maketitle

\begin{abstract}
We describe here a new method to estimate copula measure. From $N$~observations of two variables~$X$ and~$Y$, we draw a huge number~$m$ of subsamples (size~$n<N$), and we compute the joint ranks in these subsamples. Then, for $p,q\leq n$, the density in~$(p/n,q/n)$ is estimated as~$ \frac{1}{mn } \sum_{s=1}^m{\sum_{i=1}^n{\ind{R_{i, s}=p, S_{i, s}=q } } } $ where~$R_{i, s}$ (respectively~$S_{i, s}$) is the rank in~$X$ (resp.~$Y$) of the \ordinal{i}~observation of the \ordinal{s}~sample.\\
The simulation study shows that this method seems to gives a better than the usual kernel method. The main advantage of this new method is then we do not need to choose and justify the kernel. In exchange, we have to choose a subsample size: this is in fact a problem very similar to the bandwidth choice. We have then reduced the overall difficulty.

\end{abstract}

\section{Introduction}

\subsection{Copula estimation}

A first way is to estimate cumulated density function, for the bivariate observations, and for each marginal. In this way, we get Deheuvels test~\cite{deheuvels1,deheuvels2}. This estimation is consistent. Nevertheless, one state that, for example to test independence, this estimation has some drawbacks. Due to the form of the region we use to count the points, the power of the test is good if the dependence is monotonic ($\esp{X|Y=y}$ is a monotonic function of~$y$), because the regions~$\{R_i<p,S_i<q\}$ are clearly overloaded in case of an increasing dependence, and underloaded in the decreasing case. On the other hand, for a more complex dependence (for example~$Y=a \cdot X^2 +\epsilon$), the over or under loads will not be clear, on these regions.\\
A second way is to estimate copula density using kernels. In this way, we find~\cite{jdf,charp,sca}.\\
A third way is to use a parametric model, and likelihood maximum. We will not take into account this possibility in the following.

\subsection{Proposal}
The main idea is that the space of the bivariate ranks is finite: it consists in $N^2$~points for a $N$~sample. It is then possible to fill it, though avoiding smoothing. To do it, we \important{subsample}.\\
For example, if we have a sample of 30~observations, we draw many subsamples of 5~observations. Each one of these subsamples will fill 5~points out of the 25~values grid where one puts the~$R$ and the~$S$. If we draw enough subsamples, all the points will be filled out, several times if necessary, which gives a density.\\
More formally, one notes: $n$ size of the subsamples, $m$ the number of subsamples, $R_{i, s}$ (resp. $S_{i, s}$) rank of the first (resp. second) coordinate of the \ordinal{i}~observation of the \ordinal{s}~subsample and~$\delta$ the Dirac measure. The \important{probability measure} of~$(F_X(X),F_Y(Y))$ is estimated by:
	\[ \hat{\gamma} = \sum_{p,q\leq n} \left(\delta\left(\frac{p}{n},\frac{q}{n}\right) \times
	  \frac{1}{mn} \sum_{s=1}^m{\sum_{i=1}^n{\ind{R_{i, s}=p, S_{i, s}=q } } } \right)\]
An important point is that we accept to use a discretized representation of the copula. In other words, we get a density, which is continuous only with respect to a discrete measure.\\
Since we have a density, we can draw it. In the following we simulate 30~observations, with a linear correlation equal to~0.5. In the left graph, the radius of each circle is proportional to the density in each point.\\
\begin{center}\begin{tabular}{cc}
Estimated copula density & Original data \\
  \includegraphics[width=0.4\textwidth, height=0.4\textwidth]{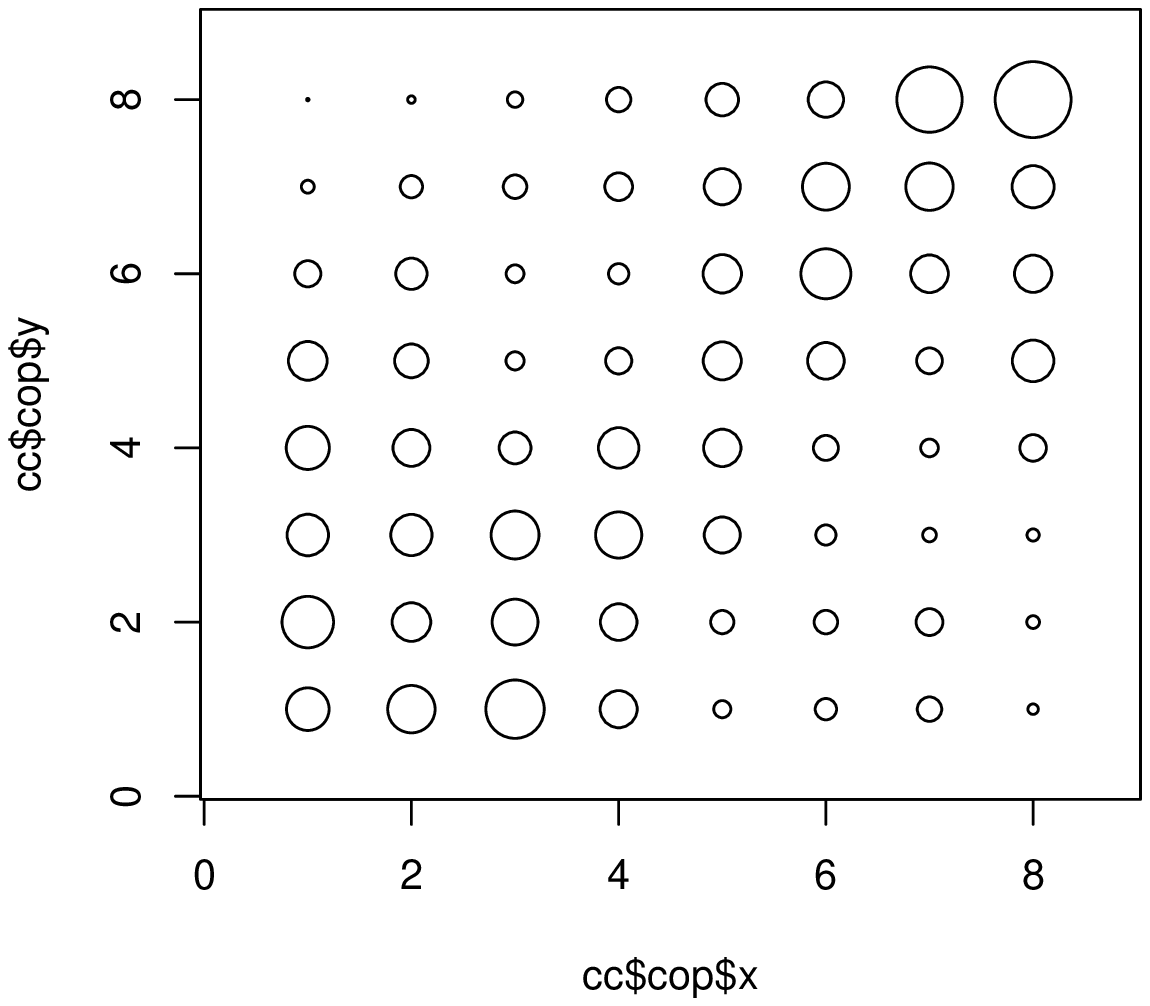} &
  \includegraphics[width=0.4\textwidth, height=0.4\textwidth]{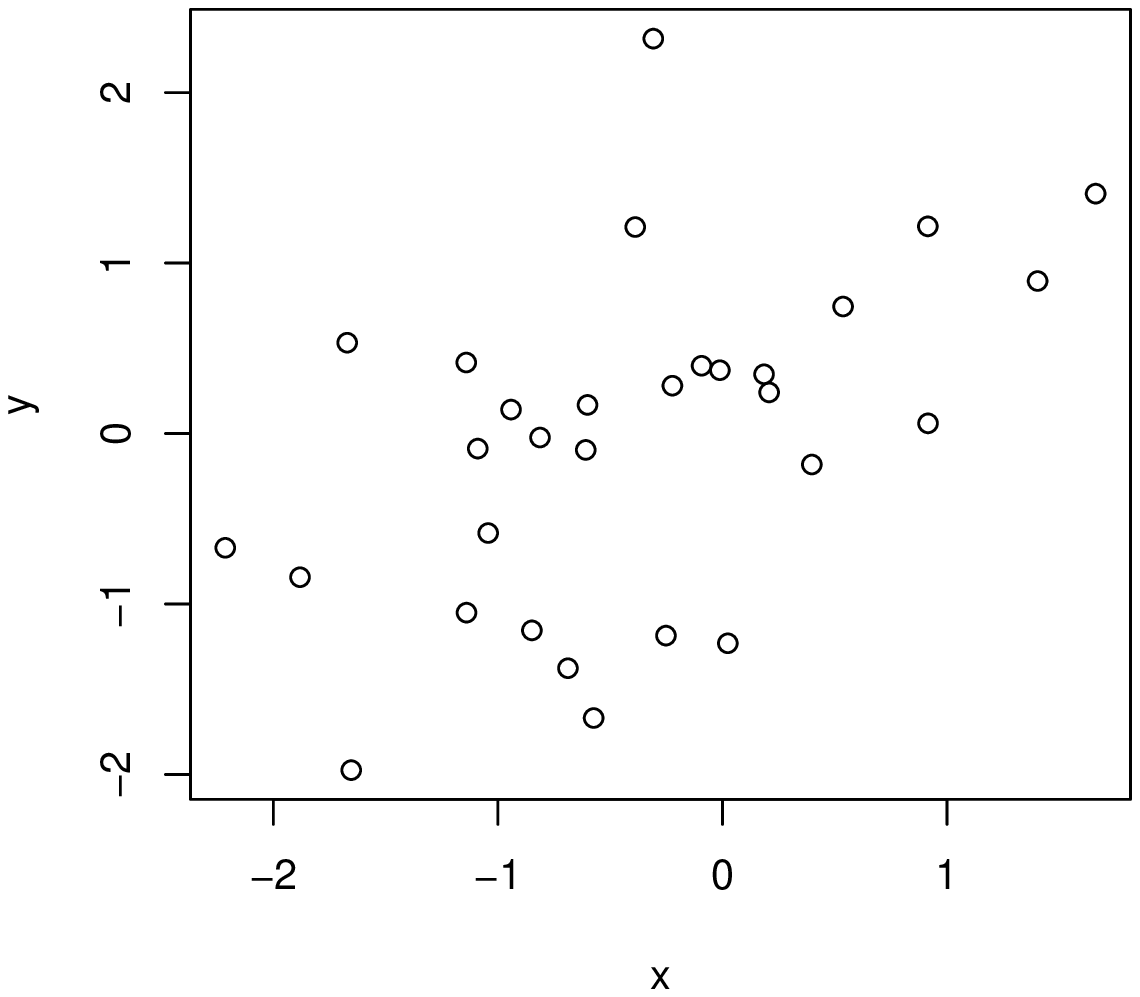}
\end{tabular}\end{center}

\paragraph{How to handle the ties?} To use this estimation in real life, we have to handle the ties. This problem, probably, has no good solution. A rank test means we assume the margins are continuous. In such a case, there would be no tie. Since we have to get rid of this problem, we will suppress the subsamples with ties. Obviously, it is possible only if the number of ties in the sample is not to big, which is necessary if we use a rank test.

\subsection{Measure convergence}
We know that, for every random variable~$X$, the empirical measure~$N^{-1} \times \sum_{i=1}^{N}{\delta(X_i)}$  converges weakly to the probability measure of~$X$. Furthermore, the empirical distribution function converges uniformly to the theoretical distribution function.\\
Then~\cite{vdw1}, the measure~$\hat{\beta}=\sum_{p,q\leq N} \left(\delta\left(\frac{p}{N},\frac{q}{N}\right) \times \frac{1}{N} \sum_{i=1}^N{\ind{R_{i}=p, S_{i}=q } }  \right)$ converges weakly to the probability measure of~$(F_X(X),F_Y(Y))$.\\
If we choose the subsample size~$n$ such that~$n^2/N \tendvers{N}{\infty} 0 $, then drawing the subsamples with or without replacement is equivalent. More precisely, the proportion of subsamples with an observation (of the sample) drawn twice tends to~0.\\
Then, for \important{each subsample}, the~$\hat{\beta}$ measure converges weakly to the same measure, which is the~$\hat{\beta}$ measure for the sample. Since~$\hat{\gamma}$ is the average of all these~$\hat{\beta}$ measures, $\hat{\gamma}$ converges weakly to the probability measure of~$(F_X(X),F_Y(Y))$.\\
Unfortunately, we need to prove the density convergence, which seems to be more difficult. Proving the convergence of a density on a finite support is tedious. One reason is that many tools used to prove convergences are designed to manage in the same way discrete and continuous issues.

\section{Goodness-of-fit test}
We follow here~\cite{sca}: this article empirically studies the power of a test based on kernel copula density estimation. To empirically study the power of a test, one simulates a lot of samples where~$H_0$ is false, and uses the test. Then, the power of the test is the proportion of samples with~$H_0$ rejected.

\subsection{Benchmark}
We study two cases.\\
In the first one, we simulate a sample with the following distribution: each observation is drawn with probability~0.5 from a Frank copula with parameter~$\theta$ , and in the remaining case from a Student copula with 4~degrees of freedom and a 0.95~correlation parameter. The test is done to decide whether this sample is drawn from a Frank copula or not. The parameter~$\theta$ has the following values: 1, 2 and~3. The Frank copula is defined by:
	\[ C(u,v)= \frac{1}{\theta} \log{\left(1+\frac{(e^{-\theta u} -1)(e^{-\theta v} -1)}{e^{-\theta} -1}\right) } \]
The second case is the same, but with the Gaussian copula replacing the Frank copula. The copula is such that, with gaussian margins, the Pearson correlation is~0.17, 0.32 or~0.47.

\subsection{Results}
We test whether a sample is drawn from \important{a} Frank copula or not. The first step is to estimate, using maximum likelihood, the parameter of a candidate Frank copula given this sample. In the following, we note this copula LFC (for Likeliest Frank Copula). Then, testing the goodness-of-fit to LFC consists in:
\begin{enumerate}
	\item\label{distance} defining and choosing a distance,
	\item\label{estnul} estimate the distribution of the distances between LFC and samples drawn from LFC,
	\item compute the distance between LFC and the original sample,
	\item reject goodness-of-fit if this distance is larger than the $(1-\alpha)$~quantile of the distribution estimated in~\ref{estnul}.
\end{enumerate}
For point~\ref{estnul}, we will use simulation (as in~\cite{sca}). The difficulty arises about point~\ref{distance} since LFC and the samples have not the same nature.\\
To be able to define a distance, we will compare the discrete densities derived from both LFC and the samples. Unfortunately, we have no numerical expression of the discrete density derived from a given theoretical copula (except for independent copula). That is why we use simulation: we compute the discrete density for a big (for example with $1000 \times N$~observations) sample drawn from LFC. Then, we only have to compare discrete densities, and we use the Kullback divergence.\\
Obviously, we need to choose a subsample size. In the following, we tried some, and chose the best one (as in~\cite{sca} for kernel bandwidth). The results are summarized in table~\ref{puissances_gof}.

\begin{table}
\begin{center}\begin{tabular}{|l|c|c|c|c|c|} \hline
Copula type & parameter & $N$ & discretized copula & subsample size & kernel density \\\hline
Gaussian & 0.17 & 50 & 0.56 & 15 & 0.39 \\
Gaussian & 0.32 & 50 & 0.38 & 13 & 0.34 \\
Gaussian & 0.47 & 50 & 0.21 & 10 & 0.21 \\\hline
Frank & 1 & 50 & 0.45 & 15 & 0.26 \\
Frank & 2 & 50 & 0.28 & 14 & 0.21 \\
Frank & 3 & 50 & 0.33 & 12 & 0.13 \\\hline
Gaussian & 0.17 & 200 & 1.00  & 13 & 1.00 \\
Gaussian & 0.32 & 200 & 0.98  & 12 & 0.99 \\
Gaussian & 0.47 & 200 & 0.86  & 11 & 0.96 \\\hline
Frank & 1 & 200 & 1.00  & 13 & 0.96 \\
Frank & 2 & 200 & 0.96  & 12 & 0.83 \\
Frank & 3 & 200 & 0.91  & 11 & 0.60 \\\hline
\end{tabular}\end{center}
\caption{\label{puissances_gof}Power with optimal choice of the subsample size}
\end{table}
Using ranks and subsampling seems to be more efficient than using kernels, at least  on this example. Furthermore, the power differences become smaller when the size of the sample increases: it would be interesting to know how would evolve the differences for smaller samples.

\section{Independence test}
\subsection{Benchmark}
We will study:
\begin{enumerate}
	\item a monotonous dependence: a simple linear relation~$y=a \cdot x + \epsilon$
	\item a non-monotonous dependence: a quadratic relation~$y=a \cdot x^2 + \epsilon$
	\item an non-functional dependence~$(x,y)= a \cdot (\cos(2\pi u),\sin(2\pi u))+(\epsilon_1,\epsilon_2)$ (called \gui{donut} in the following because of the form of the point cloud).
	\item a dependence modifying only the volatility~$y= (1+a \cdot |x|)\cdot\epsilon$ (called \gui{butterfly}).
\end{enumerate}
The variables~$\epsilon$  in dependences~1, 2, 3 and~4 and the variables~$x$ in dependences~1, 2 and~4, are normally distributed (null mean and variance unity). The variable~$u$ is uniformly distributed on~$[0,1]$ in the dependence~3.\\
We will test in the cases of~30 and 300~observations.\\
We will compare the powers of the new test, Deheuvels test, and a \gui{smart} test. This test uses an additional knowledge about the form of the dependence (but not on the value of the parameter~$a$). For the dependence~1, this test will be the Pearson test on~$y$ and~$x$, for dependence~2 on~$y$ and~$x^2$. For the dependence~3, we will test use the Komogorov-Smirnov test to know whether~$x^2+y^2$ is exponentially distributed (true if~$a=0$). For the dependence~4, we will use Pearson test on~$|x|$ and~$|y|$.\\
For Deheuvels test, we use the expression of the statistic given in~\cite{GQR}. The significativity thresholds have been computed by simulation, but they can also be checked in table~1 of~\cite{GR}.\\
The value of~$a$ will be selected so that the \gui{smart} test has a power of~0.5 or~0.9, with level~0.05. The powers are calculated on 1000~simulations.

\subsection{Results}
The test statistic is the Kullback divergence between the estimated density and the uniform density. The significativity thresholds are computed by simulation (using 3000~simulations).\\
This table~\ref{puissances_indep} gives the powers obtained for three tests: new test (column \gui{new}), Deheuvels test and \gui{smart} test. The size of the subsample is selected to maximize the power, which is obviously possible only because we have a big number of samples. In a practical setting, such a choice would be impossible.\\
So, it is obvious we need to address the subsample choice issue. We tried, \important{on the 8~dependences simulated here (for each sample size)}, a \important{minimax-regret} policy. In other words, if~$P(s,d)$ denotes the test power for dependence~$d$, and subsample size~$s$, we sought the value of~$s$ realizing:
	\[ \min_s \max_d \left[\max_s(P(s,d))-P(s,d)\right]\]
The chosen size is~8 for 30~observations and~10 for~300. The powers are in the column~\gui{new minimax} of table~\ref{puissances_indep}.\\
\begin{table}
\begin{center}\begin{tabular}{|l|c|c|c|c|c|c|c|} \hline
Dependence form & $n$ & $a$ & \gui{smart} & Deheuvels & new & optimal size & new minimax\\ \hline
Linear 			& 30  & 0.38 & 0.5 & 0.40 & 0.42 & 2 & 0.31 \\ 
Quadratic		& 30  & 0.29 & 0.5 & 0.08 & 0.21 & 10& 0.20 \\ 
Donut			& 30  & 2.90 & 0.5 & 0.04 & 0.08 & 15& 0.05 \\ 
Butterfly		& 30  & 0.87  & 0.5 & 0.07 & 0.22 & 15& 0.18 \\ \hline
Linear			& 30  & 0.67 & 0.9 & 0.82 & 0.86 & 2 & 0.76 \\ 
Quadratic		& 30  & 0.57 & 0.9 & 0.16 & 0.58 & 9 & 0.57 \\ 
Donut			& 30  & 3.76 & 0.9 & 0.04 & 0.21 & 13& 0.13 \\ 
Butterfly		& 30  & 4.9   & 0.9 & 0.11 & 0.60 & 14& 0.48 \\ \hline
Linear			& 300 & 0.11 & 0.5 & 0.50 & 0.44 & 2 & 0.34 \\ 
Quadratic		& 300 & 0.08 & 0.5 & 0.10 & 0.20 & 17& 0.17 \\ 
Donut			& 300 & 1.53 & 0.5 & 0.06 & 0.18 & 19& 0.14 \\ 
Butterfly		& 300 & 0.16 & 0.5 & 0.07 & 0.22 & 20& 0.15 \\ \hline
Linear			& 300 & 0.19 & 0.9 & 0.86 & 0.85 & 4 & 0.78 \\ 
Quadratic		& 300 & 0.14 & 0.9 & 0.20 & 0.55 & 17& 0.52 \\ 
Donut			& 300 & 1.79 & 0.9 & 0.07 & 0.40 & 19& 0.07 \\ 
Butterfly		& 300 & 0.32  & 0.9 & 0.09 & 0.55 & 21& 0.45 \\ \hline
\end{tabular}\end{center}
\caption{\label{puissances_indep}Power with optimal choice of the subsample size}
\end{table}
We have to emphasize that the relatively good power of the minimax-regret policy is not a conclusive result (oppositely, a small power would have been conclusive). It only shows that, if we are able to define it in a general and formal way, a minimax-regret policy could lead to good results.\\
One important conclusion is that the cumulated density function is not the right tool to see a non-monotonic dependence. We need something different, for example a density (even a discrete one).\\
Another important teaching (still not sure, since we did not take account of all dependencies in the minimax-regret policy) is that we can test independence, even with a very poor prior knowledge about the possible dependence. For example, the first line of the table shows that, if we assume that the dependence is linear, we are able to detect it in half of the cases. If we only have a minimax-regret policy, using density estimation, we detect it in a third of the cases. The power loss is not unbearable.

\section{Conclusion and further work}
The use of kernels for density estimation implies two choices: the kernel, the bandwidth. We have seen in this article that, in order to estimate copula measures, one can bypass the first one, using ranks and subsampling. In exchange, we have to choose a subsample size: this is in fact a problem very similar to the bandwidth choice. We have then reduced the overall difficulty.\\
The \texttt{R} and \texttt{C} code used to this simulation study is available on demand, at anyone of the addresses given in first page.\\
A theoretical article showing the convergence of the estimation is under work. Other further works would be:
\begin{itemize}
	\item studying the limiting distribution of the statistic, or at least large deviations of this statistic;
	\item studying a minimax strategy to choose the subsample size.
\end{itemize}

\end{document}